\documentclass[12pt,a4paper,twoside,dvips]{article}
\usepackage[usenames]{color}   
\usepackage{cite,mcite}
\usepackage{graphicx}
\usepackage{subscript}
\graphicspath{{./}}
\newlength{\capindent}
\newlength{\capwidth}
\newlength{\figwidth}
\newcommand{\icaption}[2][!*!,!]{\hspace*{\capindent}%
  \begin{minipage}{\capwidth}
    \ifthenelse{\equal{#1}{!*!,!}}%
      {\caption{#2}}%
      {\caption[#1]{#2}}
  \end{minipage}}
\newcommand{\eV}{\hbox{\ensuremath{\mathrm{e\kern-0.1em V}}}}%
\newcommand{\eVc}{\hbox{\ensuremath{\mathrm{e\kern-0.1em V\kern-0.16em}/\kern-0.1em\ensuremath{c}}}}%
\newcommand{\eVcsq}{\hbox{\ensuremath{\mathrm{e\kern-0.1em V\kern-0.16em}/\kern-0.1em\ensuremath{c^2}}}}%
\newcommand{\TeV}{\hbox{\ensuremath{\mathrm{T}}\kern-0.08em\eV}}%
\newcommand{\TeVc}{\hbox{\ensuremath{\mathrm{T}}\kern-0.08em\eVc}}%
\newcommand{\TeVcsq}{\hbox{\ensuremath{\mathrm{T}}\kern-0.1em\eVcsq}}%
\newcommand{\GeV}{\hbox{\ensuremath{\mathrm{G}}\eV}}%

\def\itGeV{\hbox{\it Ge\kern-1.2pt V}}%
\def\itMeV{\hbox{\it Me\kern-1.2pt V}}%
\def\itkeV{\hbox{\it ke\kern-1.2pt V}}%

\newcommand{\JETSET}{{\scshape Jetset}}

\newcommand{\PYTHIA}{{\scshape Pythia}}

\newcommand{\Lthree}{{\scshape l}{\small 3}}

\newcommand{\BE}{{\scshape BE}}

\newcommand{\BEtt}{{\BE\textsubscript{32}}}

\newcommand{\beqa}{\begin{eqnarray}}   \newcommand{\eeqa}{\end{eqnarray}}
\newcommand{\beqan}{\begin{eqnarray*}} \newcommand{\eeqan}{\end{eqnarray*}}

\newcommand{\pt}{\ensuremath{p_\mathrm{t}}}
\newcommand{\mt}{\ensuremath{m_\mathrm{t}}}
\newcommand{\mtbar}{\ensuremath{\overline{m}_\mathrm{t}}}
 
\newcommand{\Eq}[1]{Eq.\,(\ref{#1})}%
\newcommand{\Eqs}[1]{Eqs.\,(\ref{#1})}%
\newcommand{\Fig}[1]{Fig.\,\ref{#1}}%
\newcommand{\ee}{\ensuremath{\mathrm{e^{+}e^{-}}}}
\newcommand{\eg}{\textit{e.g.}}%
\newcommand{\ie}{\textit{i.e.}}%

\newcommand{\Letter}{article}

\begin{document}
\title{Parametrization of Bose-Einstein Correlations \\
        and Reconstruction of the Space-Time Evolution of Pion Production\\
       in e$^+$e$^-$ Annihilation}
 
\author{
   T. Cs\"org\H{o}
  \\ \textit{\small MTA KFKI RMKI, H-1525 Budapest 114, Hungary}    \\[2mm]
   W. Kittel,
   W.J. Metzger,
   T. Nov\'ak\footnotemark[3]
  \\ \textit{\small Radboud University, NL-6525 AJ\ \ Nijmegen, The Netherlands}
}
%

 
\footnotetext[3]{Present address: Dept.\ of Business Mathematics, K\'aroly R\'obert College, H-3200 Gy\"ongy\"os, Hungary}
\maketitle

\begin{abstract}
A parametrization of the Bose-Einstein correlation function of pairs of identical pions
produced in hadronic \ee\ annihilation is proposed within the framework of a model
(the $\tau$-model) in which space-time and momentum space are very strongly correlated.
Using information from the Bose-Einstein correlations as well as from single-pion spectra,
it is then possible to reconstruct the space-time evolution of pion production.
\end{abstract}

\section{Introduction}\label{sect:intr}
In particle and nuclear physics, intensity interferometry provides a direct
experimental method for the determination of sizes, shapes and lifetimes
of particle-emitting sources
(for reviews see, \eg, \cite{Gyulassy:1979,Boal:1990,Baym:1998,Wolfram:Zako2001,Tamas:HIP2002}).
In particular, boson interferometry provides a powerful tool for the
investigation of the space-time structure of particle production processes,
since Bose-Einstein correlations (BEC) of two identical bosons reflect both
geometrical and dynamical properties of the particle radiating source.
Given the point-like nature of the underlying interaction, \ee\ annihilation provides an ideal environment
to study these properties in multiparticle production by quark fragmentation.
 
\section{Bose-Einstein Correlation Function}
The two-particle correlation function of two particles with
four-momenta $p_{1}$ and $p_{2}$ is given by the ratio of the two-particle number density,
$\rho_2(p_{1},p_{2})$,
to the product of the two single-particle number densities, $\rho_1 (p_{1})\rho_1 (p_{2})$.
Being only interested in the correlation $R_2$ due to Bose-Einstein
interference, the product of single-particle densities is replaced by
$\rho_0(p_1,p_2)$,
the two-particle density that would occur in the absence of Bose-Einstein correlations:
\begin{equation} \label{eq:R2def}
  R_2(p_1,p_2)=\frac{\rho_2(p_1,p_2)}{\rho_0(p_1,p_2)} \;.
\end{equation}
Since the mass of the two identical particles of the pair is fixed to the pion mass,
the correlation function is defined in six-dimensional momentum space.
Since Bose-Einstein correlations can be large only at small four-momentum difference
$Q=\sqrt{-(p_1-p_2)^2}$,
they are often parametrized in terms of this one-dimensional distance measure.
There is no reason, however,
to expect the hadron source for jet fragmentation to be spherically symmetric.
Recent investigations, using the Bertsch-Pratt parametrization~\cite{Pratt:86,Bertsch:88},
have, in fact, found an elongation of the source along the
jet axis~\cite{L3_3D:1999,OPAL3D:2000,DELPHI2D:2000,ALEPH:2004,OPAL:2007}
in the longitudinal center-of-mass (LCMS) frame~\cite{tamas:workshop91}.
While this effect is      well established, the elongation is actually only about 20\%,
which suggests that a parametrization in terms of the single variable $Q$,
may be a good approximation.
 
There have been indications that the size of the source, as measured using BEC, depends on the transverse
mass, $\mt=\sqrt{m^2+\pt^2}=\sqrt{E^2-p_z^2}$, of the pions~\cite{Smirnova:Nijm96,Dalen:Maha98,OPAL:2007}.
It has been shown~\cite{Bialas:1999,Bialas:2000} that such a dependence can be understood if the produced
pions satisfy, approximately, the (generalized) Bjorken-Gottfried
condition~\cite{Gottfried:1972,Bjorken:SLAC73,Bjorken:1973,Gottfried:1974,Low:1978,Bjorken:ISMD94}, whereby
the four-momentum of a produced particle and the space-time position at which it is produced are linearly
related:   $ x     = d p     $.
Such a correlation between space-time and momentum-energy is  also a feature of the Lund string model
as incorporated in \JETSET\ \cite{JETSET74},
which is very successful in describing detailed features of the hadronic final states of \ee\ annihilation.
Recently, experimental support for this strong correlation has been found \cite{OPAL:2007}.
 
A model which predicts both a $Q$- and an \mt-dependence while incorporating the  Bjorken-Gottfried
condition is the so-called $\tau$-model \cite{Tamas;Zimanji:1990}.
In this \Letter\ we develop this model further and apply it to the reconstruction of
the space-time evolution of pion production in \ee\ annihilation.

\section{BEC in the \boldmath{$\tau$} model}  \label{sect:taumodel}
In the $\tau$-model, it is assumed that the average production point in the overall center-of-mass system,
$\overline{x}=(\overline{t},\overline{r}_x,\overline{r}_y,\overline{r}_z)$, of particles with a given
four-momentum $p$ is given by
\begin{equation} \label{eq:tau-corr}
   \overline{x}     (p)  = a\tau p     \;.
\end{equation}
In the case of two-jet events,
$a=1/\mt$
where
\mt\ is the transverse mass
and
$\tau = \sqrt{\overline{t}^2 - \overline{r}_z^2}$ is the longitudinal proper time.\footnote{The
terminology `longitudinal' proper time and `transverse' mass seems customary in the literature
even though their definitions are analogous $\tau = \sqrt{\overline{t}^2 - \overline{r}_z^2}$
and                                         $ \mt = \sqrt{E^2            - p_z^2}$.}
For isotropically distributed particle production, the transverse mass is replaced by the
mass in the definition of $a$ and $\tau$ is the proper time.  
In the case of three-jet events the relation is more complicated.
 
The correlation between coordinate space and momentum space variables
is described by the distribution of $x     (p)$ about its average by
$\delta_\Delta ( x    (p    ) - \overline{x}(p ) ) = \delta_\Delta(x-a\tau p)$.
The emission function of the $\tau$-model is then given by \cite{Tamas;Zimanji:1990}
\begin{equation}  \label{eq:source}
  S(x,p) = \int_0^{\infty} \mathrm{d}\tau H(\tau)\delta_{\Delta}(x-a\tau p) \rho_1(p) \;,
\end{equation}
where $H(\tau)$ is the (longitudinal) proper-time distribution
and $\rho_1(p)$ is the experimentally measurable single-particle momentum spectrum,
both $H(\tau)$ and $\rho_1(p)$ being normalized to unity.
 
The two-pion distribution, $\rho_2(p_1,p_2)$, is related to $S(x,p)$, in the plane-wave approximation,
by the Yano-Koonin formula~\cite{Yano}:
\begin{equation}  \label{eq:yano}
   \rho_2(p_1,p_2) = \int \mathrm{d}^4 x_1 \mathrm{d}^4 x_2 S(x_1,p_1) S(x_2,p_2) 
                     \left\{\strut 1 + \cos\left[ (p_1-p_2)(x_1-x_2) \right]\strut\right\} \;.
\end{equation}
Assuming that the distribution of $x (p)$ about its average
is much narrower than the proper-time distribution,
\Eq{eq:yano} can be evaluated in a saddle-point approximation.
Approximating
the function $\delta_\Delta$                     by a Dirac delta function yields the same result.
Thus the integral of  \Eq{eq:source} becomes
\begin{equation}  \label{eq:S}
                   \int_0^{\infty} \mathrm{d}\tau H(\tau) \rho_1\left(\frac{x}{a\tau}\right) \;,
\end{equation}
and the argument of the cosine in \Eq{eq:yano} becomes
\begin{equation}   \label{eq:cos}
   (p_1 - p_2)(\bar{x}_1 - \bar{x}_2) = - 0.5 (a_1\tau_1 + a_2\tau_2) Q^2 \;.
\end{equation}
Substituting \Eqs{eq:S} and (\ref{eq:cos}) in \Eq{eq:yano} leads to the following approximation of
the two-particle Bose-Einstein correlation function:
\begin{equation} \label{eq:levyR2}
   R_2(Q,a_1,a_2) = 1 + \mathrm{Re} \widetilde{H}\left(\frac{a_1 Q^2}{2}\right)
                                    \widetilde{H}\left(\frac{a_2 Q^2}{2}\right) \;,
\end{equation}
where $\widetilde{H}(\omega) = \int \mathrm{d} \tau H(\tau) \exp(i \omega \tau)$
is the Fourier transform of $H(\tau)$.
 
This formula simplifies further if $R_2$ is measured with the restriction
\begin{equation} \label{eq:a1a2equal}
     a_1\approx a_2\approx \bar{a} \;.
\end{equation}
In that case, $R_2$ becomes
\begin{equation}     \label{eq:levyR2a}
   R_2(Q,\bar{a}) = 1 + \mathrm{Re} \widetilde{H}^2 \left(  \frac{\bar{a}Q^2}{2} \right)
   \;.
\end{equation}
Thus for a given average of $a$ of the two particles, $R_2$  is found to
depend only on the invariant relative momentum $Q$.
Further, the model predicts a specific dependence on $\bar{a}$, which for two-jet events is
a specific dependence on  $\overline{m}_\mathrm{t}$.\footnote{In the initial formulation of
the $\tau$-model
this dependence was averaged over \cite{Tamas;Zimanji:1990} due to the lack of \mt\ dependent data
at that time.}

Since there is no particle production before the onset of the collision,
$H(\tau)$ should be a  one-sided distribution.
We choose a one-sided L\'evy distribution, which has the characteristic function (Fourier transform)~\cite{Tamas:Levy2004}
(for $\alpha\ne1$)\footnote{For the special case $\alpha=1$, see, \eg, Ref.~\citen{Nolan}.}
\begin{equation} \label{eq:levy1sidecharf}
   \widetilde{H}(\omega) = \exp\left\{ -\frac{1}{2}\left(\Delta\tau|\omega|\strut\right)^{\alpha\strut}
          \left[ 1 -  i\, \mathrm{sign}(\omega)\tan\left(\frac{\alpha\pi}{2}\right) \strut \right]
       + i\,\omega\tau_0\right\}                          \;,
\end{equation}
where the parameter $\tau_0$ is the proper time of the onset of particle production
and $\Delta \tau$ is a measure of the width of the proper-time distribution.
Using this characteristic function in \Eq{eq:levyR2a} yields
\begin{eqnarray} \label{eq:levyR2av}
   R_2(Q,\bar{a}) &=&
       1+ \cos \left[\strut{\bar{a}\tau_0 Q^2}
                                   + \tan \left( \frac{\alpha \pi}{2} \right)
                                     \left( \frac{\bar{a}\Delta\tau Q^2}{2}\right)^{\!\alpha\strut} \right]
           \nonumber \\
   & & \cdot            \exp \left[\strut -\left( \frac{\bar{a}\Delta\tau Q^2}{2}\right)^{\!\alpha\strut} \right] 
        \;,
\end{eqnarray}
which for two-jet events is
\begin{eqnarray} \label{eq:levy2jetR2av}
   R_2(Q,\mtbar) &=&
       1+ \cos \left[ \frac{\tau_0 Q^2}{\mtbar}
                                   + \tan \left( \frac{\alpha \pi}{2} \right)
                                     \left( \frac{\Delta\tau Q^2}{2\mtbar}\right)^{\!\alpha\strut} \right]
           \nonumber \\
   & & \cdot            \exp \left[ -\left( \frac{\Delta\tau Q^2}{2\mtbar}\right)^{\!\alpha\strut} \right]
        \;.
\end{eqnarray}

\begin{figure}[htb]
  \centering
  \includegraphics*[width=.95\figwidth]{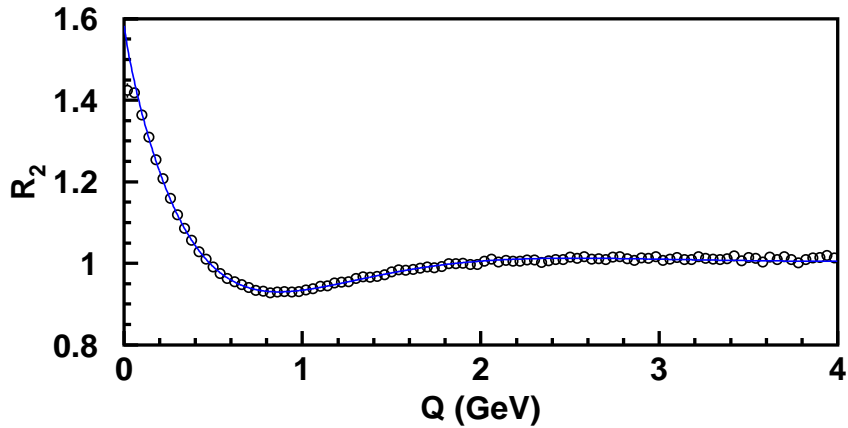}
  \caption{The Bose-Einstein correlation function $R_2$ for events generated by \PYTHIA.
      The curve corresponds to a fit of the one-sided L\'evy parametrization, \Eq{eq:asymlevR2}.
           \label{fig:a_levy}
           }
\end{figure}

We now   consider a simplification of the equation
obtained by assuming (a) that particle production starts immediately, \ie, $\tau_0=0$,
and (b) an average $a$-dependence, which is implemented in an approximate way by defining an effective
radius, $R=\sqrt{\bar{a}\Delta\tau/2}$, which for 2-jet events becomes
$R=\sqrt{\Delta\tau/(2\overline{m}_\mathrm{t})}$.
This results in:
\begin{equation}\label{eq:asymlevR2}
    R_2(Q) = 1+ \cos \left[(R_\mathrm{a}Q)^{2\alpha}\right] \exp \left[-(RQ)^{2\alpha} \right]      \;,
\end{equation}
where $R_\mathrm{a}$ is  related to $R$ by
\begin{equation}\label{eq:asymlevRaR}
    R_\mathrm{a}^{2\alpha} = \tan\left(\frac{\alpha\pi}{2}\right) R^{2\alpha} \;.
\end{equation}
To illustrate that \Eq{eq:asymlevR2} can provide a reasonable parametrization, we show in \Fig{fig:a_levy}
a fit of \Eq{eq:asymlevR2} with $R_\mathrm{a}$ a free parameter
to Z-boson decays generated by \PYTHIA\ \cite{PYTHIAsix}
with BEC simulated by the \BEtt\ algorithm \cite{LS98} as tuned to \Lthree\ data \cite{L3:QCDphysrep}.
In particular, it describes well the dip in $R_2$ below unity in the $Q$-region 0.5--1.5\,\GeV, unlike
the usual Gaussian or exponential parametrizations.
While generalizations \cite{tamas:edge:lag00}
of the Gaussian  by  an  Edgeworth expansion and of the exponential  by  a Laguerre
expansion can describe the dip, they require more additional parameters than
\Eq{eq:asymlevR2}.
Recently the \Lthree\ Collaboration has presented preliminary results showing that \Eq{eq:asymlevR2}
describes their data on hadronic Z decay \cite{wes:WPCF2006}.
 
\section{The emission function of two-jet events}  \label{sect:emission2jet}
Within the framework of the $\tau$-model, we now show how to
reconstruct the space-time picture of pion emission.
We restrict ourselves to two-jet events where we know what $a$ is, namely $a=1/\mt$.
The emission function in configuration space, $S_\mathrm{x}(x)$, is the proper time derivative of the
integral over $p$ of $S(x,p)$, which in the
$\tau$-model is given by \Eq{eq:source}.
Approximating $\delta_\Delta$ by a Dirac delta function, we find
\begin{equation}   \label{eq:Sspace}
   S_\mathrm{x}(x) = \frac{1}{\bar{n}} \frac{\mathrm{d}^4 n}{\mathrm{d}\tau\mathrm{d}^{3}x}
                   = \left(\frac{\mt}{\tau}\right)^3 H(\tau) \rho_1\left( p=\frac{\mt x}{\tau} \right) \;,
\end{equation}
where $n$ and $\bar{n}$ are the number and average number of pions produced, respectively.
 
Given the symmetry of two-jet events, $S_\mathrm{x}$ does not depend on the azimuthal angle, and we can
write it in cylindrical coordinates as
\begin{equation}   \label{eq:Srzt}
     S_\mathrm{x}(r,z,t) = P(r,\eta) H(\tau) \;,
\end{equation}
where $\eta$ is the space-time rapidity.
With the strongly correlated phase-space of the $\tau$-model,
$\eta=y$ and $r=\pt\tau/\mt$.
Consequently,
\begin{equation}  \label{eq:Preta}
     P(r,\eta) = \left(\frac{\mt}{\tau}\right)^{\!3} \rho_\mathrm{\pt,y}(r\mt/\tau, \eta) \;,
\end{equation}
where $\rho_\mathrm{\pt,y}$ is the joint single-particle distribution of \pt\ and $y$.
 
The reconstruction of $S_\mathrm{x}$ is simplified if $\rho_\mathrm{\pt,y}$
can be factorized in the product of the single-particle \pt\ and rapidity distributions, \ie,
$     \rho_\mathrm{\pt,y} = \rho_\mathrm{\pt}(\pt) \rho_\mathrm{y}(y)$.
Then \Eq{eq:Preta} becomes
\begin{equation}   \label{eq:fact}
     P(r,\eta) = \left(\frac{\mt}{\tau}\right)^{\!3} \rho_\mathrm{\pt}(r\mt/\tau) \rho_\mathrm{y}(\eta) \;,
\end{equation}


 
The transverse part of the emission function is obtained by integrating over $z$ as well as azimuthal angle.
Pictures of this function evaluated at successive times would together form a movie revealing the time
evolution of particle production in 2-jet events in \ee\ annihilation.
 

To summarize:  Within the $\tau$-model, $H(\tau)$ is obtained from a fit of
\Eq{eq:levy2jetR2av}  to the Bose-Einstein correlation function.
From $H(\tau)$ together with the inclusive distribution of rapidity and \pt,
the full emission function in configuration space, $S_\mathrm{x}$, can then be reconstructed.

%
\section*{Acknowledgments}
One of us (T.C.) 
acknowledges support of
the Scientific Exchange between Hungary (OTKA) and The Netherlands (NWO),
project B64-27/N25186 as well as Hungarian OTKA grants T49466 and NK73143.
 
%
 
\bibliographystyle{l3style}
\bibliography{%
generators,%
jets,%
bec,levy%
}

\end{document}